\documentclass{article}
\usepackage[utf8]{inputenc} 
\usepackage[T1]{fontenc}    
\usepackage[total={6in, 8in}]{geometry}
\usepackage{hyperref}       
\usepackage{url}            
\usepackage{booktabs}       
\usepackage{amsfonts}       
\usepackage{amsthm}
\usepackage{nicefrac}       
\usepackage{microtype}      
\usepackage{xcolor}         
\usepackage{graphicx}
\usepackage{float} 
\usepackage{mathtools}
\usepackage{placeins}
\usepackage{natbib}
\usepackage{tabularx}
\usepackage{wrapfig}
\usepackage{multicol}
\usepackage{authblk} 

\title{Amazon Ads\\Multi-Touch Attribution\thanks{Correspondence: Randall Lewis, causal@amazon.com; Florian Zettelmeyer, flozet@amazon.com.}}

\author[1]{Randall Lewis}
\author[1,2,3]{Florian Zettelmeyer}
\author[1,2]{Brett R.~Gordon}
\author[1]{Cristobal Garib}
\author[4]{Johannes Hermle}
\author[4]{Mike Perry}
\author[1]{Henrique Romero}
\author[1]{German Schnaidt}
\affil[1]{Amazon}
\affil[2]{Northwestern University}
\affil[3]{NBER}
\affil[4]{Work done at Amazon}

\makeatletter

\date{August 11, 2025}

\begin{document}
\maketitle

\begin{abstract}
Amazon’s new Multi-Touch Attribution (MTA) solution allows advertisers to measure how each touchpoint across the marketing funnel contributes to a conversion. This gives advertisers a more comprehensive view of their Amazon Ads performance across objectives when multiple ads influence shopping decisions. Amazon MTA uses a combination of randomized controlled trials (RCTs) and machine learning (ML) models to allocate credit for Amazon conversions across Amazon Ads touchpoints in proportion to their value—i.e., their likely contribution to shopping decisions. ML models trained purely on observational data are easy to scale and can yield precise predictions, but the models might produce biased estimates of ad effects. RCTs yield unbiased ad effects but can be noisy. Our MTA methodology combines experiments, ML models, and Amazon’s shopping signals in a thoughtful manner to inform attribution credit allocation.
\end{abstract}

\section{What is the advertiser’s problem?}

Today, customers' journeys are no longer linear and involve multiple touchpoints across the entire marketing funnel. An advertiser’s goal is to drive customer conversions to maximize the overall return on investment (ROI) of their ad budget across marketing channels. Learning how each channel contributes to business outcomes helps advertisers assess the effectiveness of their marketing spend.
 
The key question is ``how much does each touchpoint contribute to a conversion?'' For instance, how much, if at all, do earlier upper-funnel touchpoints contribute to an eventual conversion through a lower-funnel marketing tactic?
 
Striking the right balance of investments across channels is critical. Advertisers need to invest enough in upper-funnel tactics to support brand awareness and consideration. Yet they also need sufficient investments in lower-funnel tactics to win over those customers already considering their brands. Advertisers’ challenge is to balance marketing budgets across channels based on their relative effectiveness in order to implement their preferred full-funnel marketing strategy.

\section{Assessing Attribution}

Last-touch attribution (LTA) credits the full value of a conversion to one ad, usually the last one on the path to purchase, overlooking the impact of earlier touchpoints like awareness campaigns that build interest before shoppers are ready to buy. This approach can lead advertisers to undervalue and reduce investment in tactics that play a key role in conversions. To address this point, industry-standard Multi-Touch Attribution (MTA) is a measurement approach that tracks all touchpoints along the customer journey before a conversion, assigning specific credit to each touchpoint. The purpose of these credits is to help advertisers understand which tactics work, so they can decide which to reinforce, change, or abandon. 

Consider an advertiser that spent \$1,000 each on upper- and lower-funnel campaigns on different marketing channels (Figure 1). The upper-funnel campaign sought to drive brand awareness, increasing sales by \$2,000---sales that would not have happened without the campaign (i.e., incremental sales). 
\begin{wrapfigure}{r}{0.49\textwidth}
\centering
\includegraphics[width=0.49\textwidth]{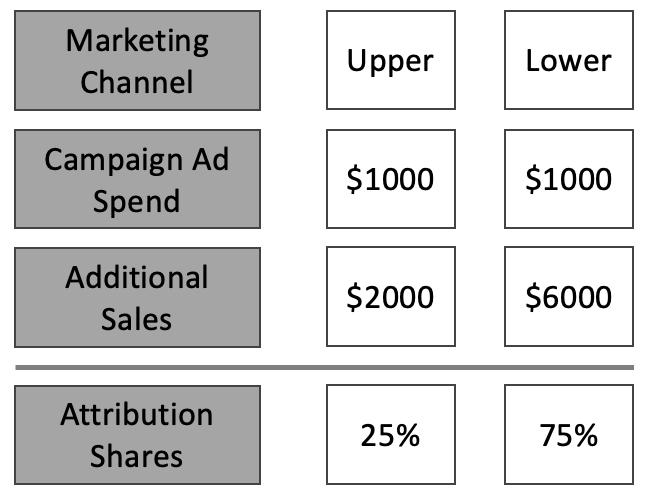}
\caption{Attribution Shares \label{attribution_shares}}
\end{wrapfigure}
However, the lower-funnel campaign, which promoted specific products, increased sales by \$6,000. To help the advertiser learn which tactic worked better, an attribution algorithm should reveal that the product-targeting campaign was more effective than the brand awareness campaign and, ideally, quantify that it was three times as effective. The \textbf{Attribution Shares} represent this relative effectiveness. Therefore, the best attribution algorithm is the one that most accurately identifies the \emph{relative} incremental effect of the advertiser’s choices on their conversion outcomes. Such an attribution algorithm solves the advertiser’s problem: generating credits proportional to incremental effects helps them reallocate ad spending to tactics that yield higher returns. Do industry-standard MTA approaches achieve this goal?

Common MTA algorithms use preexisting rules to allocate credit to different touchpoints. The simplest MTA algorithm is a linear model that places equal weight on all touchpoints in the conversion path. However, attributing equal weight to every touchpoint oversimplifies the process, as it disregards crucial factors such as context and timing, thereby failing to accurately reflect the value generated by each advertisement. To address this limitation, some advertisers employ a more nuanced approach, adjusting the weights based on the sequence or timing of the touchpoints. One popular method is to use an ``exponential decay'' model that assigns more weight to touchpoints that occur closer to the conversion. This method requires the advertiser to make several choices: what precise decay rate should be used? Should the decay differ by ad product or product category? How should the method handle clicks versus views? 

Newer MTA approaches incorporate machine learning models trained on non-experimental ``observational'' data (i.e., data collected in the incidental course of business). These algorithms can be used to determine custom weights for each touchpoint. Such models have the advantage of incorporating customer signals from many sources while leveraging a robust approach to prediction evaluation developed in computer science. Still, advertisers face various choices when relying on such methods: which features should they include? Which machine learning model is the right one? And most importantly, how should we judge whether any of these decisions are ``accurate''?

A general problem with the MTA methods discussed earlier is that they lack a comparison with the incrementality of a campaign. Without such a ``ground truth,'' it is difficult to evaluate any particular MTA solution. Thus, we need a reliable way to measure a campaign’s incrementality in order to determine the attribution shares on any given tactic.

When machine learning models only rely on observational data---lacking ground truth---research from both industry and academia has shown that they are unable to accurately and reliably estimate the incremental impact of ads. For example, a 2023 study of almost 2,000 representative campaigns at Meta showed that sophisticated models using observational data led to errors in the estimated ad effects ranging from 488\% to 948\%, depending on the model and outcome type \citep{gordon2023close}. Relying purely on observational data too often produces biased estimates of incremental ad effects, making such an approach unreliable for an attribution system. 

\section{Getting Ground Truth}

To address the advertiser's measurement problem, we need a way to understand the incremental value of each touchpoint for customers. Specifically, how much less likely would customers have been to convert if they hadn't seen a particular ad? Quantifying the impact of individual touchpoints is key to optimizing overall ad spend.

\subsection{A useful thought experiment}

To accurately assess the impact of an advertisement, we would ideally like to compare a customer's behavior in two parallel scenarios that are identical in every respect, except for one key difference: in one scenario, the consumer is exposed to the advertisement, while in the other, they are not. These two hypothetical ``worlds'' should be exactly the same, with the sole difference being the presence or absence of the advertisement. If such a controlled comparison were feasible, and we observed a difference in outcomes (e.g., purchases) between the two scenarios, we could attribute that difference to the incremental effect of the advertisement, because all other factors would be held constant.
 
Although this thought experiment is conceptually useful, the fundamental challenge in establishing causality lies in the fact that a consumer cannot simultaneously exist in two parallel worlds---it is impossible for an individual to both view and not view an advertisement at the exact same moment.

\subsection{The solution: RCTs}

The recognized solution to this problem is to run a true experiment, or what is commonly referred to as a \textbf{Randomized Controlled Trial} (RCT). An RCT is a robust method used across many fields, including medicine, social science, and marketing, to evaluate the effectiveness of interventions. The underlying principle is to randomly assign participants to one of several distinct conditions or ``worlds'' and then compare outcomes across these conditions to measure the effect of an intervention. In an advertising RCT, one group of shoppers is eligible to see ads from the focal campaign (``treatment'') while another group of shoppers is not (``holdout''). Using Amazon's shopping signals, we compare outcomes between these two groups to measure the causal effect of the ad campaign on actual shoppers' behavior. 
 
Even if a large number of consumers, say 100,000 or more, are randomly divided into treatment and holdout, the resulting groups may not be precisely identical. Each group inevitably comprises different individuals. Does this mean we cannot truly isolate the effect of the intervention from potential differences between the two groups? 
 
The answer is to realize that randomization creates ``probabilistically equivalent'' groups, allowing us to compare outcomes as if consumers were in parallel worlds, even though we can't observe the same individual in both conditions simultaneously. The power of randomization lies in its ability to balance all consumer characteristics, including \emph{unobserved} factors (e.g., latent brand awareness or purchase intent), across the treatment and holdout groups. This eliminates systematic differences in characteristics or ad responsiveness across the groups. With large, truly randomized groups, any difference in outcomes between conditions can be attributed to the ad itself, not to differences in consumer characteristics.
 
RCTs provides a clear measure of incremental impact by evaluating the effect of the marketing activity (treatment group) against a baseline where no intervention occurs (holdout group). This contrasts with other types of tests that compare two or more marketing strategies without a holdout group. It is noteworthy that the Interactive Advertising Bureau (IAB) and Media Rating Council (MRC) released updated Retail Media Measurement Guidelines in 2024 that ``highly recommends using RCTs'' to measure advertising incrementality \citep{iab2024retail}. 
 
Despite their strengths, RCTs have two significant limitations. First, RCTs can produce results that are too imprecise (i.e., wide confidence intervals) to yield conclusive answers about ad performance. Second, RCTs generate ad effects at the level of a campaign, but an MTA model require credits assigned at each touchpoint. 
 
Thus, although RCTs provide ground truth measures of incrementality, they are too noisy and too coarse to serve alone as a solution to the MTA problem. In principle, we could run RCTs at massive scale with large holdout groups (e.g., ``test all ad lines with 50\% of customers in the control group'') to ameliorate the coarseness and noise problem. However, doing so would deprive advertisers of a large fraction of their desired audience and be very costly. 
 
Instead, our solution will use a \emph{combination of RCTs and machine learning (ML) based attribution models} to allocate credit for Amazon conversions across Amazon Ads touchpoints. As we describe in the next section, ML-based attribution models trained purely on observational data are easy to scale and can yield precise predictions, but the models might produce biased estimates of ad effects. RCTs yield unbiased ad effects but can be noisy. The MTA approach we describe in section 5 combines experiments, ML models, and Amazon’s shopping signals in a thoughtful manner to inform the allocation of credit.

\section{The Value of Information within Attribution Models}

Relative to RCTs, attribution models approach ad measurement from a completely different perspective. RCTs start by randomizing treatment (e.g., ad vs. no ad) and then measure the difference in outcomes (e.g., purchases) between treatment and holdout groups. In contrast, an attribution model starts with an \emph{outcome} and then divides credit for this outcome among preceding touchpoints (e.g., views, clicks, etc.). Last-touch attribution (LTA) gives 100\% of the credit to the touchpoint that most closely precedes the outcome. More sophisticated attribution models use machine learning methods and rich sets of features to determine each touchpoint's credit share.
 
One key distinction between causal effect estimation and attribution models is that attribution models use information that occurred \emph{after the treatment} to assign credit. For example, the treatment in an ad campaign is the ad impression. Attribution models typically use information on whether users clicked on an ad (which happened after the treatment, i.e., the ad impression) as diagnostic information in assigning credit to the ad for those users. Clearly, such ``post-determined'' (post-treatment) information helps us assign credit for a conversion to different ads because some were clicked—while others were not—and this likely informs us about the effect of each ad. The consequence is that we can use clicks and other post-determined information to flesh out the customer journey triggered by the marketing interventions. If our hypotheses about the journey are correct, such models improve the accuracy and precision of attributed ad effects. 
 
RCTs cannot use post-determined information such as ad clicks because they require users in the treatment and holdout groups to be comparable. If we attempt to measure the ad effect for users who clicked on the ad, the problem is that clicking is only possible for users who were exposed to the ad (in the treatment group). Consequently, we cannot identify comparable users in the holdout group, as they were never given the opportunity to click. This lack of equivalence between the groups undermines our ability to draw valid causal conclusions by using post-determined information.
 
This discussion highlights the core dilemma we aim to solve using Amazon’s MTA methodology: RCT-based ad effects provide ground truth but cannot use any post-determined information (like a click). This makes RCT effects noisy for assigning credit to touchpoints (like a click) because most touchpoints \emph{are} post-determined information! Attribution models \emph{do} use post-determined information, which allows us to assign credit to touchpoints and increase the precision of attributed ad effects. However, as we noted earlier, other research has shown that attributed ad effects, which use post-determined information, are likely biased relative to the RCT ad effects.

We want to combine the best of both worlds: to use RCTs because they provide unbiased measures of ad effects and to incorporate the post-determined information from attribution models to make the results more precise. This is the approach introduced in Predicted Incrementality by Experimentation (PIE) \citep{gordon2025predictive}. Our extension also produces touchpoint-level attribution credits. 

\section{Combining RCTs and Attribution Models for Better Measurement}

To give the intuition for how Amazon MTA works, we begin by showing how to \emph{calibrate an attribution model using RCTs}. This corresponds to building a simple prediction model where we use the attribution model’s attributions as a feature to predict the treatment effects from RCTs. This works because a portion of attributed conversions are typically incremental, which implies a partly mechanical relationship between the attributed and RCT conversions. Next, we show how to extend the prediction model to use the attributions from multiple attribution models as features, i.e., we construct an ensemble of attribution models. So far, the unit of analysis has been an RCT, an aggregation of consumers into treatment or holdout groups for some ad campaign. Finally, we move from a prediction model at the campaign level to scoring individual touchpoints, allowing us to later aggregate touchpoint-level credits into attribution shares to inform the advertiser’s problem.

\subsection{Calibration Factor Models}

We focus on a single upper-funnel marketing channel, called ``Upper.'' Suppose an RCT for Upper tells us the campaign generated 900 incremental conversions, whereas the 7-day last-touch attributed (LTA) conversions (based on relevance rules) comes in at 1,000. A simple \textbf{calibration factor model} would be:
\begin{align} \label{eq1}
\text{RCT Conversions}_{\text{Upper}} & = \beta \times \text{LTA Conversions}_{\text{Upper}} \ . 
\end{align}
This RCT implies that the calibration factor is $\beta = 0.9$. The interpretation of this calibration factor is that there are 10\% fewer incremental conversions than what we think when using a last-touch attributed metric. This calibration factor could be applied to future ad campaigns \emph{that are not implemented as RCTs} as a way to calibrate, or adjust, the LTA conversions to obtain a prediction of the RCT conversions for that campaign. 

The concept behind this calibration factor model may already be familiar to many advertisers, with some potentially using similar approaches. However, this example relies on data from a single RCT for one ad campaign. If an advertiser were to conduct multiple RCT campaigns, a more robust model, such as linear regression, could be used to better understand the relationship between RCT and LTA conversions. The resulting value of $\beta$ would represent the average calibration factor across these campaigns for the advertiser, providing a more comprehensive assessment. 
 
Advertisers often face limitations in conducting enough RCTs to implement such a robust model effectively. However, at Amazon, we have the advantage of access to extensive signals, which have allowed us to significantly expand upon and refine this basic idea.

\subsection{Ensemble of Attribution Models}

The calibration factor model is a simple \textbf{predictive model} that combines RCTs effects (the target) and an attribution model (the feature). One basic strategy to improve any prediction model is to obtain more features that are (hopefully) predictive of the outcome variable. In our setting, the additional features take the form of \emph{other attribution models}.
 
So far, we have considered one feature, namely the number of LTA conversions. In addition to LTA, we could incorporate a linear attribution model that assigns equal credit across all touchpoints. This formula for assigning credit can improve our predictive accuracy because it would incorporate more information about the customer's journey, for example, prior views or clicks on Amazon, instead of focusing only on the last touchpoint as with LTA. Still, both LTA and linear attribution represent rule-based approaches to credit assignment.

At Amazon, rather than using a rule to assign credit, we can use much more sophisticated attribution models that use machine learning and rich customer signals to determine attribution credit. These model-driven attributions---``MDA,'' for short---may be better at predicting the RCT ad effects because they can more flexibly extract the right relationships from the historical data in ways that rule-based attribution models cannot. Therefore, we include MDA conversions for predicting the RCT outcomes with the overall goal of improving the model's predictive performance.
 
Doing so expands the model to an ensemble of attribution models: 
\begin{align} \label{eq2}
\text{RCT Conversions}_{c} & = f\left(\text{LTA Conversions}_{c},\  \text{MDA Conversions}_{c}\right) \ .
\end{align}
To illustrate how this might work in a simple example, imagine we have a collection campaigns for an advertiser that were each run as RCTs on upper-funnel (``Upper'') and lower-funnel (``Lower'') marketing channels. We also have the corresponding LTA conversions and MDA conversions for each campaign, such that we can extend our calibration factor model to a multi-channel setting:
\begin{align} \label{eq3}
\text{RCT Conversions}_{c} & = \beta \times \text{LTA Conversions}_{c} \ + \ \alpha \times \text{MDA Conversions}_{c} \ ,
\end{align}
where $c=\{\mathrm{Upper},\mathrm{Lower}\}$. After implementing the model, we obtain estimates of $\beta=0.6$ and $\alpha=0.4$. These estimates allow us to predict the RCT conversions for each marketing channel even for campaigns for which we do not have RCTs.

We refer to the above as a \textbf{Causal Calibration Model}, which serve two purposes: (1) they help extend the RCT ad effects to other ad campaigns with similar features, even those that were not run as RCTs and (2) they allow us to generate MTA credits for ad campaigns at any level of granularity.

\subsection{From Campaigns to Touchpoints}

We use the campaign-level Causal Calibration Model to generate predictions at the touchpoint level. Mechanically, this is straightforward, because the features for the campaign-level model originated at the touchpoint level and were subsequently aggregated to features at the campaign level. 
 
The overall process contains four steps. First, we train the campaign-level model in Equation 2 (Equation 3 in this specific simple example) and evaluate how well it predicts out-of-sample RCT outcomes. The attribution models that serve as features (i.e., LTA, MDA) rely on touchpoint-level signals aggregated up to the campaign level. Second, we disaggregate these features---our collection of attribution models' attributed conversions---from the campaign to the touchpoint level. Third, we use these values to generate touchpoint-level predictions of ad effects, which we call ``MTA credits.'' Fourth, once we have these MTA credits, we can aggregate and normalize them for advertiser-facing reporting, just as we already do with LTA credits. For example, perhaps an advertiser is interested in comparing performance across different types of ad campaigns.

To illustrate, imagine that Amazon had run two marketing campaigns, one in marketing channel ``Upper'' and one in marketing channel ``Lower.'' Both campaigns were run as RCTs. We make use of two attribution models, namely LTA and MDA, to predict RCT outcomes. Using the model in Equation 3 to determine the weights for LTA and MDA attributions, we find that $\beta=0.6$ and $\alpha=0.4$. This means we should multiply LTA-attributed conversions by $\beta=0.6$ and multiply MDA-attributed conversions by $\alpha=0.4$, and then add up the weighted conversions from both models to best predict the RCT conversions.

\begin{figure}[H]
\centering
\includegraphics[width=0.95\textwidth]{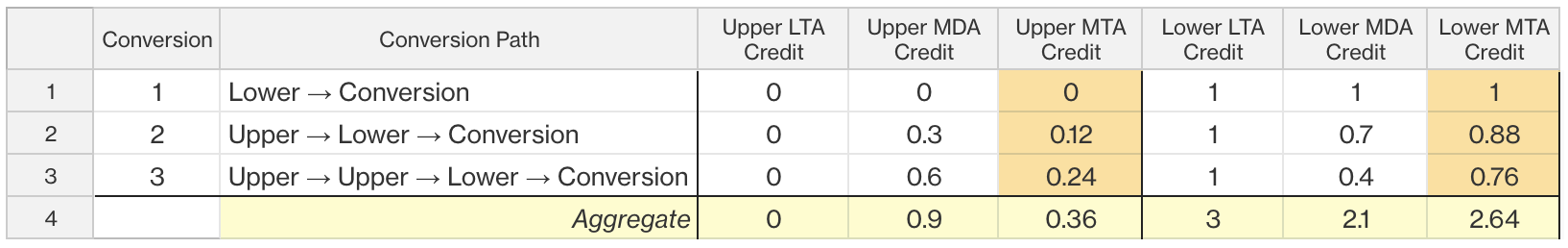}
\caption{Example of Credit Assignment at the Touchpoint Level. \label{attribution_ex1}}
\end{figure}

Now consider three \textbf{converting customers} who were touched by either Upper, Lower, or both channels as part of these campaigns, as in Figure~\ref{attribution_ex1}. LTA assigns full credit (a value of one) to the last touchpoint. MDA was built to be more flexible, and therefore assigns credit between zero and one to each touchpoint. For example, the full credit for customer 2's conversion is given to the Lower channel under LTA. Under MDA, however, both Upper and Lower channels receive credit (0.3 and 0.7, respectively) since both channels touched customer 2 once.

Instead of using LTA or MDA credit, we use MTA credit, which simply multiplies each LTA credit by 0.6 and each MDA credit by 0.4. The MTA credits (in orange above) represent the model’s predictions of the effect of each touchpoint on each conversion.
 
\begin{figure}[H]
\centering
\includegraphics[width=0.95\textwidth]{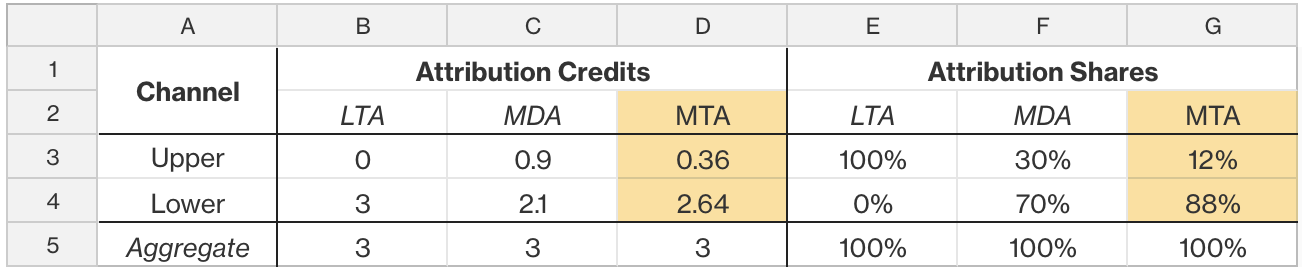}
\caption{Example of Attribution Credits and Shares at the Channel Level.\label{attribution_ex2}}
\end{figure}

If we were only interested in these three conversions, we could aggregate the results in the previous table (see yellow-shaded row of Figure~\ref{attribution_ex1}) to the level of each marketing channel. Figure~\ref{attribution_ex2} on the next page presents Attribution Credits aggregated over the three conversions and then normalized into Attribution Shares. The Attribution Shares for each ad product are proportional to the value they generate for these conversions. As one can see, the MTA shares differ from both LTA and MDA because they are informed by RCTs. 

The MTA Credits reflect a weighted average of the credit assigned from either of the two attribution models. The weights are determined by (a) the existence of campaign-level RCTs and (b) the model's ability to accurately predict the RCT results. Thus, this approach represents a principled way to combine the unbiased results from RCTs with the additional signals found in attribution models, whether these attribution models are rule-based (like LTA) or model-informed (like MDA). Of course, the example above is simple compared to the myriad of conversion paths experienced by our customers: they might have multiple marketing interactions across a variety of channels at different points in time. 

We hope you now have some intuition for the underlying mechanism of this approach to MTA. Next, we describe more about what happens to make the system work at Amazon scale.

\section{Amazon MTA}

Amazon's new MTA solution allows advertisers to measure how each touchpoint across the marketing funnel contributes to a conversion. This gives advertisers a more comprehensive view of their Amazon Advertising performance across objectives when multiple ads contribute to a conversion.
 
Amazon MTA uses a combination of RCTs and machine learning (ML) models to allocate credit for Amazon conversions across Amazon Ads touchpoints in proportion to their value—i.e., their likely contribution to shopping decisions. ML models trained purely on observational data are easy to scale and can yield precise predictions, but the models might produce biased estimates of ad effects. RCTs yield unbiased ad effects but can be noisy. Our MTA methodology combines experiments, ML models, and Amazon's shopping signals in a thoughtful manner to inform the allocation of credit. 

\begin{figure}[t]
\centering
\includegraphics[width=0.95\textwidth]{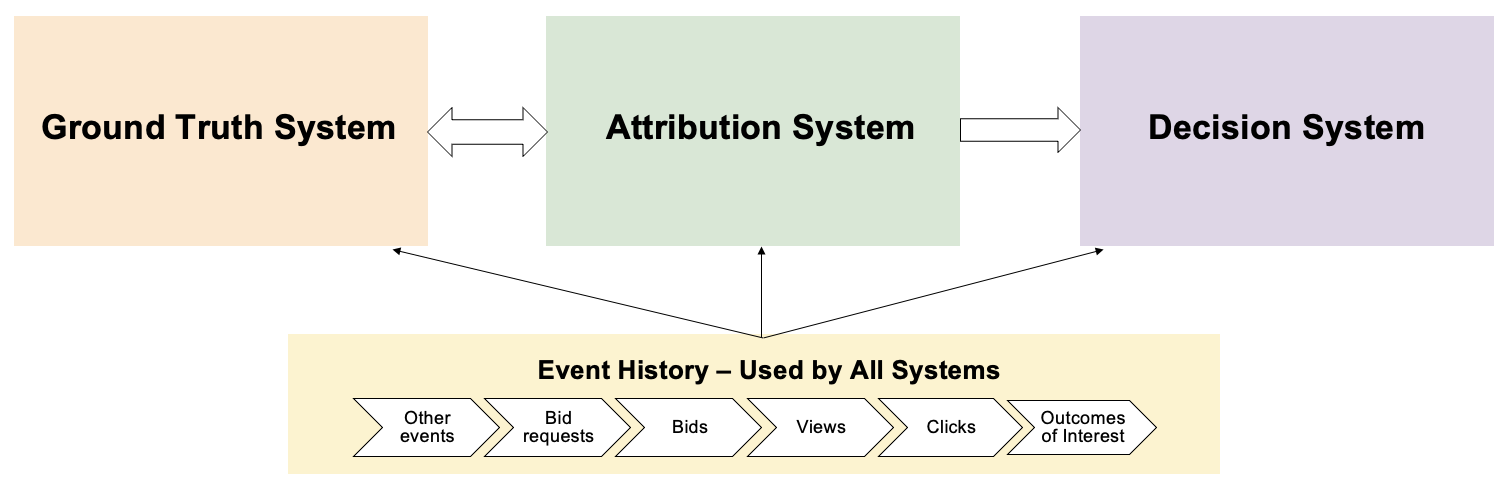}
\caption{Amazon Multi-Touch Attribution (MTA) System.\label{mta_system}}
\end{figure}

Amazon MTA is comprised of three systems, which we briefly describe in Figure~\ref{mta_system}. First, the Ground Truth System contains the collection of RCTs and the Causal Calibration Models. Second, following a conversion event, the Attribution System generates attribution credits based on the Causal Calibration Models and translates them into attribution shares. Third, the Decision System incorporates attribution scores into optimization models. Each system relies on Event History data that merges signals from a variety of sources.

\subsection*{Ground Truth System}

At Amazon, we have a representative sample of RCTs across our ad products that provides training data with ground truth causal signals of campaign effectiveness. 
 
The quality of the MTA system depends on maintaining a substantial, representative, and current database of RCTs. We follow a deliberate approach to generate and maintain this dataset. The strategy we adopt to determine which campaigns to run as RCTs helps ensure we obtain RCTs on a representative collection of ad campaigns for each ad product, while minimizing opportunity costs. Over time we adjust this strategy based on MTA system performance and to reflect changes in the broader advertising industry. The result is a database of hundreds of thousands of RCTs across our ad products that spans most of our advertisers and campaign types.
 
These training data feed the Causal Calibration Models that help us extend the RCT ad effects in the training data to other ad campaigns with similar features, even for campaigns that were not run as RCTs. 

\subsection*{Attribution System}

The Attribution System serves two main purposes. First, it provides attribution models that serve as features for the Causal Calibration Models in the Ground Truth System. These models are designed to be flexible because advertisers face different marketing situations across ad campaigns and products. For example, our ensemble of attribution models includes a causal machine learning model that assesses the impact of each marketing touchpoint by comparing sales outcomes in the presence and absence of ads. This model captures signals from a variety of sources, including traffic events, bid logs, retail information, ASIN catalogs, and more. It is based on a deep learning framework designed to be highly flexible and to allow for potential interactions across input signals and outcomes. As the business environment changes and customers evolves, we can further augment the ensemble of attribution models to incorporate new models and signals to more accurately predict RCT outcomes.
 
Second, the Attribution System takes as input the Event History and the trained Causal Calibration Models to generate attribution credits for each type of touchpoint associated with a conversion event. These credits are indexed by a set of characteristics (e.g., ad product, delivery channel, placement type, etc.) and only touchpoints within the allowed lookback window are considered. These touchpoint-level credits are aggregated into attribution credits based on the relevant set of touchpoints, given the particular characteristics of each touchpoint. The result is a set of causally-defensible multi-touch attribution shares that can inform the advertiser’s problem.

\subsection*{Decision System}

The attribution credits are the focal input for the Decision System. Ad products use attribution credits, which describe ad effectiveness, to train optimization models to help improve the effectiveness of advertisers’ ad spend.

\bibliographystyle{apalike} 

\bibliography{mta}

\end{document}